\documentclass[twocolumn,aps,prd,amsmath,amssymb,nofootinbib]{revtex4-1}\newcommand{\preprintnumber}{\hfill MIT-CTP 4356\,\,\,\,\maketitle}
\usepackage{graphicx,bm}

\newcommand{\beq}{\begin{equation}}
\newcommand{\bea}{\begin{eqnarray}}
\newcommand{\eeq}{\end{equation}}
\newcommand{\eea}{\end{eqnarray}}

\newcommand{\mpl}{M_{Pl}}

\begin{document}

\title{Inflation, Symmetry, and B-Modes}

\author{Mark P.~Hertzberg}
\affiliation{Center for Theoretical Physics and Dept.~of Physics,\\ Massachusetts Institute of Technology, Cambridge, MA 02139, USA}

\date{\today}

\begin{abstract}
We examine the role of using symmetry and effective field theory in
inflationary model building. We describe the standard formulation of
starting with an approximate shift symmetry for a scalar field, and
then introducing corrections systematically in order to maintain
control over the inflationary potential. We find that this leads to
models in good agreement with recent data. On the other hand, there are
attempts in the literature to deviate from this paradigm by invoking
other symmetries and corrections. In particular: in a suite of recent
papers, several authors have made the claim that standard Einstein
gravity with a cosmological constant and a massless scalar carries
conformal symmetry. They claim this conformal symmetry is \emph{hidden}
when the action is written in the Einstein frame, and so has not been
fully appreciated in the literature. They further claim that such a
theory carries another hidden symmetry; a global $SO(1,1)$ symmetry. By
deforming around the global $SO(1,1)$ symmetry, they are able to
produce a range of inflationary models with asymptotically flat
potentials, whose flatness is claimed to be protected by these
symmetries. These models tend to give rise to B-modes with small
amplitude. Here we explain that standard Einstein gravity does not in
fact possess conformal symmetry. Instead these authors are merely
introducing a \emph{redundancy} into the description, not an actual
conformal symmetry. Furthermore, we explain that the only real (global)
symmetry in these models is not at all hidden, but is completely
manifest when expressed in the Einstein frame; it is in fact the
\emph{shift symmetry} of a scalar field. When analyzed systematically
as an effective field theory, deformations do not generally produce
asymptotically flat potentials and small B-modes as suggested in these
recent papers. Instead, deforming around the shift symmetry
systematically, tends to produce models of inflation with B-modes of appreciable
amplitude. Such simple
models typically also produce the observed red spectral index, Gaussian
fluctuations, etc. In short: simple models of inflation, organized by
expanding around a shift symmetry, are in excellent agreement with
recent data.
\end{abstract}

\preprintnumber

%\newpage
%\tableofcontents

%s1 ###
%s1 #&#
\section{Introduction}

\let\thefootnote\relax\footnotetext{Electronic address: {\tt mphertz@mit.edu}}

The theory of cosmological inflation \cite{Guth,Linde}, a phase of
acceleration expansion in the early universe, is in good agreement with
a range of observations. It is able to account for the large-scale
homogeneity and isotropy of the universe, as well as providing a
beautiful mechanism for the origin of large scale fluctuations. A
missing component of the theory is a preferred model for the
inflationary dynamics, although many interesting models have been
proposed throughout the last few decades.

The simplest inflationary models involve Einstein gravity sour\-ced by a
scalar field $\phi$ and a potential $V(\phi)$. If we truncate the
action at two derivatives, the action can be written, without loss of
generality, as
%
%e1 ###
%e1 #&#
%
\begin{eqnarray}
S=\int\! d^4x\sqrt{-g}\,\left[ {\mpl^2\over2}R+{1\over2}(\partial
\phi)^2-V(\phi)\right]
\end{eqnarray}
where $\mpl\equiv1/\sqrt{8\pi G}$ and we are using the signature
$+--\,-$. If we choose the potential to simply be a cosmological
constant, we would have a possibility of de Sitter space, though it
would never end. So one normally imagines that the potential has some
shape to it, including a minimum with $V\sim0$, where inflation can
end. The slow-roll conditions for a prolonged phase of inflation are
%
%e2 ###
%e2 #&#
%
\begin{eqnarray}
\epsilon\equiv{\mpl^2\over2}\left({V'\over V}\right)^2; \,\,
\epsilon\ll1,\nonumber\\
\eta\equiv\mpl^2\left({V''\over V}\right); \,\,|\eta|\ll1.
\label{slowroll}
\end{eqnarray}
These conditions typically require the potential to be rather flat over
a Planckian or super-Planckian domain in field space $\Delta\phi$. A~generic potential $V(\phi)$ would not usually have this property. In
fact many generic potentials that emerge in top-down models do not have
this property. For example, if we parameterize the potential as a
series expansion in powers of $\phi$ as follows (let's impose a $\phi
\to-\phi$ symmetry for simplicity)
%
%e3 ###
%e3 #&#
%
\begin{eqnarray}
V(\phi)=\Lambda_0+{1\over2}m^2\phi^2+{\lambda\over4}\phi^4+\sum
_{n=6}^\infty{c_n\over\mpl^{n-4}}\phi^n
\end{eqnarray}
Then if the coefficients are some fairly random numbers, and lets say
$c_n\sim\mathcal{O}(1)$, then the required flatness of the potential
is usually spoiled.

So to make progress one often invokes some type of symmetry structure.
The most basic version of this is to imagine that $\phi$ carries a
\emph{shift symmetry} $\phi\to\phi+\phi_0$. This sets all the
coefficients of the above potential to zero and obviously leaves a flat
potential. But since this would be too strong, one then relaxes the
shift symmetry slightly, i.e., allows a weak breaking of the shift
symmetry by introducing very small values for the $c_n$, etc. This is
said to be ``technically natural'' as the symmetry is restored in the
limit in which the coefficients are set to zero. As we will describe
later, this idea ultimately underpins the ``chaotic inflation'' model
\cite{Linde2}.
Related arguments occur for ``natural inflation'' in which one imagines
that $\phi$ is a Goldstone boson associated with some spontaneously
broken (global) symmetry \cite{Adams:1992bn}. This automatically
forces the coefficients $c_n$ to vanish. One then imagines that the
underlying global symmetry is broken by some quantum effects, perhaps
by non-perturbative effects as in the case of the axion, to generate
small but non-zero coefficients.
In other contexts, such as string theory, other possible structures
emerge to control symmetries, such as ``monodromies'', which can control
the shape of the potential in an interesting way \cite
{Silverstein:2008sg,McAllister:2008hb}. We will carefully study this
general framework in Section~\ref{EFT}.

Currently, we do not know if any of these symmetry arguments are on the
right track, but they do organize the action into a sensible effective
field theory and lead to interesting testable predictions. With
fantastic precision in recent CMB observations \cite
{Hinshaw:2012aka,Ade:2013uln,Ade:2014xna}, including polarization data,
this program of model building is very worthwhile.

In this paper, we examine a recent claim of a new class of inflation
models based on conformal symmetry. In Section~\ref{New} we describe
these models. In Section~\ref{WhatConf} we review the meaning of
conformal symmetry. In Section~\ref{WhatConfNot} we explain why this
new class of models does not carry a physical conformal symmetry. In
Section~\ref{Global} we make the actual physical (global) symmetry in
the models manifest and recognize it as a standard shift symmetry. In
Section~\ref{EFT} we show how to deform around this standard shift
symmetry within the framework of effective field theory. In Section~\ref
{Bmode} we discuss the consequences of various models for the
amplitude of B-modes; contrasting those based on fine tuning and those
based on symmetry. Finally, we discuss in Section~\ref{Discuss}.

%s2 ###
%s2 #&#
\section{New class of symmetry models?}\label{New}

Recently, a new class of inflation models organized by symmetry was put
forward by several authors \cite
{Kallosh:2013pby,Kallosh:2013lkr,Kallosh:2013hoa,Kallosh:2013maa,Kallosh:2013daa,Kallosh:2013yoa,Kallosh:2013oma,Linde:2014nna}
(related ideas are also being examined in the context of bouncing
cosmologies \cite{Bars:2011aa,Bars:2013yba,Bars:2013vba}). The basic
new claim centers around the structure of standard Einstein gravity. It
is claimed that standard Einstein gravity, even with a cosmological
constant, carries a \emph{conformal symmetry}. This is quite a
dramatic claim, especially since such a model appears to carry two mass
scales: the Planck mass $\mpl$ and the energy scale of the
cosmological constant $\Lambda^{1/4}$. If we include a massless scalar
field, the action is the following:
%
%e4 ###
%e4 #&#
%
\begin{eqnarray}
S=\int\! d^4x\sqrt{-g}\,\left[ {\mpl^2\over2}R+{1\over2}(\partial
\phi)^2-\Lambda\right]
\label{GFaction}
\end{eqnarray}
So how could it possibly be that such a theory is actually conformal?
The answer, they say, is that this conformal symmetry is \emph{hidden}
\cite
{Kallosh:2013pby,Kallosh:2013lkr,Kallosh:2013hoa,Kallosh:2013maa,Kallosh:2013daa,Kallosh:2013yoa,Kallosh:2013oma,Linde:2014nna}.

To exhibit this hidden conformal symmetry they introduce another scalar
field $\chi$ which forms a doublet with the other scalar under a
global $SO(1,1)$ symmetry, as follows \cite{Linde:2014nna}
%
%e5 ###
%e5 #&#
%
\begin{eqnarray}
S&=&\int\! d^4x\sqrt{-g}\,\Bigg[ {1\over12}(\chi^2-\psi^2)R+{1\over
2}(\partial\psi)^2\nonumber\\
&&{}
\,\,\,\,\,\,\,\,\,\,\,\,
-{1\over2}(\partial\chi)^2-{\lambda\over4}(\psi^2-\chi^2)^2\Bigg]
\label{GIaction}
\end{eqnarray}
Notice that the kinetic term for $\chi$ is negative; which is a ghost
term (however, in this context it does not lead to an instability).
This action does not contain any explicit dimensionful parameters. The
dramatic claim is that this action is \emph{conformal} and it is
connected to the above action. To claim this, the authors point out
that this action is unchanged under the following transformations
%
%e6 ###
%e6 #&#
%
\begin{eqnarray}
g_{\mu\nu}&\to& e^{2\alpha}g_{\mu\nu}\nonumber\\
\psi&\to& e^{-\alpha}\psi\nonumber\\
\chi&\to& e^{-\alpha}\chi
\label{GT}
\end{eqnarray}
which they refer to as a ``local conformal symmetry''. Since $\alpha$
is an arbitrary function we can use it to ``gauge fix'' the scalar
fields in a way we choose. In particular we can gauge fix
%
%e7 ###
%e7 #&#
%
\begin{eqnarray}
\chi^2-\psi^2= 6\mpl^2
\end{eqnarray}
This condition can be parameterized by writing
%
%e8 ###
%e8 #&#
%
\begin{eqnarray}
\psi&=& \sqrt{6}\,\mpl\sinh(\phi/\sqrt{6}\,\mpl)\nonumber
\\
\chi&=& \sqrt{6}\,\mpl\cosh(\phi/\sqrt{6}\,\mpl)
\end{eqnarray}
Then upon substitution into Eq.~(\ref{GIaction}) we find the action
given in Eq.~(\ref{GFaction}) of standard Einstein gravity
with a free
massless scalar and a cosmological constant $\Lambda=9\lambda\mpl^4$.
So it would appear as though standard Einstein gravity with a
cosmological constant is actually conformally invariant, but that its
conformal symmetry is hidden by gauge fixing.

The next step is to deform the symmetries in order to build interesting
models for inflation. The procedure that has been advocated is to
return to the action in Eq.~(\ref{GIaction}) and keep the conformal
symmetry intact (they say it is a local or gauge symmetry so it should
not be broken), but they choose to break the global $SO(1,1)$ symmetry
in the following way \cite{Linde:2014nna}
%
%e9 ###
%e9 #&#
%
\begin{eqnarray}
S&=&\int\! d^4x\sqrt{-g}\,\Bigg[ {1\over12}(\chi^2-\psi^2)R+{1\over
2}(\partial\psi)^2\nonumber\\
&&{}
\,\,\,\,\,\,\,\,\,
-{1\over2}(\partial\chi)^2-{\lambda\over4}F(\psi/\chi)(\psi
^2-\chi^2)^2\Bigg]
\label{GIactionDeform}
\end{eqnarray}
where $F$ is some dimensionless function of the ratio of $\psi$ to
$\chi$. Notice that this action is unchanged under the transformations
given in Eq.~(\ref{GT}) although $F$ breaks the global $SO(1,1)$
symmetry (unless $F$ is a constant). Then by gauge fixing to the
Einstein frame, as before, we are led to the following gauge fixed action
%
%e10 ###
%e10 #&#
%
\begin{eqnarray}
S=\int\! d^4x\sqrt{-g}\,\left[ {\mpl^2\over2}R+{1\over2}(\partial
\phi)^2-V(\phi)\right]
\label{GFactionDeform}
\end{eqnarray}
with
%
%e11 ###
%e11 #&#
%
\begin{eqnarray}
V(\phi) = \Lambda\, F(\operatorname{tanh}(\phi/\sqrt{6}\,\mpl)
\label{Vtanh}
\end{eqnarray}
(again with $\Lambda=9\lambda\mpl^4$). This has the nice property
that for many choices of $F$, such as $F(x)\propto x^{n}$, this
potential $V(\phi)$ asymptotes to a constant at large
(super-Planckian) field values. Since it asymptotes to a constant for
super-Planckian field values then we can expect slow-roll inflation to
occur at such values. Indeed, the slow-roll conditions $\epsilon\ll1$
and $|\eta|\ll1$ will be readily satisfied for many choices of $F$.
So it is quite impressive by simply appealing to some symmetries, in
particular a conformal symmetry and a deformed global symmetry, one can
build many models of slow-roll inflation with asymptotically flat potentials.
One also finds that these models generally predict \cite{Kallosh:2013daa}
%
%e12 ###
%e12 #&#
%
\begin{eqnarray}
n_s\approx1-{2\over N_e},\,\,\,\,\,r\approx{12\over N_e^2}
\end{eqnarray}
where $n_s$ is the scalar spectral index, $r$ is the tensor-to-scalar
ratio, and $N_e$ is the number of e-foldings of inflation (usually
$50\lesssim N_e\lesssim60$). We will discuss these predictions further
in Section~\ref{Bmode}.

In the rest of this note, we show that while these are some beautiful
ideas, the above analysis hides some important subtleties. In
particular, (i) by carefully defining conformal symmetry, we show that
these models \emph{do not} actually carry conformal symmetry, and (ii)
by deforming around the global symmetry in the sense of effective field
theory, we show that these models do not generically yield
asymptotically flat potentials. We also comment on some other
interesting attempts in the literature to obtain a conformal theory of
gravitation.

%s3 ###
%s3 #&#
\section{What conformal symmetry is}\label{WhatConf}

Let us begin by defining conformal symmetry in the context of field
theory. The first ingredients we need are some matter degrees of
freedom $\psi_i$, and some dynamics governed by a Lagrangian $\mathcal
{L}$. Lets allow for some non-trivial metric $g_{\mu\nu}$ that is
treated as a \emph{background}. The action is
%
%e13 ###
%e13 #&#
%
\begin{eqnarray}
S=\int\! d^4x\sqrt{-g}\,\mathcal{L}(\psi_i,\partial_\mu\psi_i)
\end{eqnarray}
The idea is to ask the following question: Does the action change if we
perform a conformal change to the metric? That is, if we consider a
background metric $g_{\mu\nu}$ and then rescale it as follows
%
%e14 ###
%e14 #&#
%
\begin{eqnarray}
g_{\mu\nu}\to\Omega(x)^2g_{\mu\nu}
\label{weyl}
\end{eqnarray}
we wish to know if the dynamics is different in this new metric. Notice
that the idea is to really change the actual metric, not simply our
representation of the metric, i.e., we wish to explore different
space--times, not a mere rewriting of a given same space--time.

We may also allow the $\psi_i$ to transform with some power of $\Omega
$ as
%
%e15 ###
%e15 #&#
%
\begin{eqnarray}
\psi_i \to\Omega(x)^{\Delta_i} \psi_i
\label{scalingdimension}
\end{eqnarray}
where $\Delta_i$ is known as the ``scaling dimension'' of $\psi_i$. If
for some choice of $\Delta_i$ the action returns to itself, then we
obviously have a symmetry, a so-called ``conformal symmetry''. In this
special circumstance the physics is unchanged for different choices of
conformally related metrics.

Some simple examples include pure electromagnetism, $\mathcal{N}=4$
super-Yang Mills, and massless $\lambda\phi^4$ theory with
non-minimal coupling to the background Ricci scalar $-\phi^2 R/12$.
The first two of these examples are exact at the quantum level, while
the third example is only true classically. One consequence of the
conformal symmetry is that the trace of the stress-energy tensor vanishes.
Notice that it obviously requires a very special form for the
Lagrangian for this conformal symmetry to exist. For instance, the
Lagrangian obviously cannot possess any explicitly dimensionful
parameters, such as mass terms, as this would immediately violate scale
invariance (which is a necessary condition for conformal invariance).

%s4 ###
%s4 #&#
\section{What conformal symmetry is not}\label{WhatConfNot}

%s4.1 ###
%s4.1 #&#
\subsection{Dynamical space--time}

In the previous section we defined a conformal symmetry for some matter
degrees of freedom with respect to some \emph{background} metric.
Could it be possible that a conformal symmetry can extend to the case
of a dynamical metric? Indeed, the claim of these authors is that the
action given in Eq.~(\ref{GIactionDeform}) is conformally invariant
when treating \emph{both} the scalar (matter) fields as dynamical and
the metric itself as dynamical.

Indeed, it is true that for the action given in Eq.~(\ref{GIaction}),
it is unchanged after performing the transformation of Eqs.~(\ref{weyl}) and (\ref{scalingdimension}) with $\Delta
_i=-1$ for the pair
of scalar fields; this was earlier described in Eq.~(\ref{GT}) with
$\Omega=e^\alpha$. However, there is a very important difference
between the case of a background metric and a dynamical metric. In the
case of a background metric the transformation in Eq.~(\ref{weyl})
changes the actual metric. However in the case of a dynamical metric
this transformation is actually just a \emph{field redefinition}. This
does not change the actual metric, but only the \emph{representation}
of the metric. This is actually true for any gauge transformation; they
leave the fields/states \emph{invariant}, by definition.

Hence the transformations reported earlier in Eq.~(\ref{GT}) are
merely gauge transformations and not an actually changing of the metric.
This is associated with the fact that there is a \emph{redundant}
degree of freedom in the action. This redundancy can by eliminated by
gauge fixing. We did this earlier; we cut down from two scalar fields
to one, by gauge fixing to the so-called Einstein frame.

Real symmetries are precisely those that remain \emph{after} gauge
fixing.$^1$\footnote{$^1$ In some cases, the symmetries can be hidden
after gauge fixing. For example, the Higgs mechanism can hide internal
(global) symmetries when we gauge fix in the unitary gauge. However,
even in this case, the symmetry is still manifest in some sectors of
the theory and the global symmetry can still be checked to be present
by the identification of a conserved quantity by the Noether theorem.
In the models studied here, there are no sectors of the theory that
carry the purported conformal symmetry, nor any conserved quantities.
On the other hand, there is a real global $SO(1,1)$ symmetry,
which is, indeed, manifest after gauge fixing;
we will return to this in Section~\ref{Global}.}
In this case it is simple to see that the theory does not have a
conformal symmetry, since the Einstein frame gauge fixed action shows
that there exist explicit mass scales that break scale (and conformal)
symmetry; namely the Planck mass $\mpl$ and the energy scale of the
cosmological constant $\Lambda^{1/4}$. Furthermore, it is relatively
straightforward to see that there are loop corrections that generate a
tower of higher dimension (derivative) operators, suppressed by the
Planck scale. This evidently breaks conformal symmetry. Also, if we
examine the deformed action expressed in the Einstein frame (see
Eqs.~(\ref{GFactionDeform}, \ref{Vtanh}) the
existence of the
potential shows that conformal symmetry is broken. For instance, a
$\lambda\phi^4$ term carries a conformal anomaly, etc. Furthermore,
for a potential of the form $V\sim\operatorname{tanh}(\phi/\sqrt{6}\,
\mpl
)$, we can Taylor expand it around $\phi=0$, and see that it is
evidently a tower of operators which, even at the classical level,
break conformal symmetry.

Instead for a theory of gravitation to carry conformal symmetry, when
gravity is treated dynamically, requires some very special structure; a
point we will return to in Section~\ref{ConfGravity}.

%s4.2 ###
%s4.2 #&#
\subsection{Background space--time}

To further drive home this point, lets turn to another case where it is
extremely important to disentangle field redefinitions from actual
field changes. This problem can even emerge when studying a fixed
background space--time.

To begin, consider the following action of a single scalar field $\phi
$ without dynamical gravity. We may in fact be simply interested in
flat space, or conformally flat space, but lets include a metric to
express the action in a generally co-ordinate invariant way
%
%e16 ###
%e16 #&#
%
\begin{eqnarray}
S &=& \int\! d^4 x\sqrt{-g}\Bigg[{1\over2}g^{\mu\nu}\partial_\mu
\phi\partial_\nu\phi\nonumber\\
&&{}
-{1\over2}m^2\phi^2-{\lambda\over4}\phi^4-\sum_{n=6}^\infty
{c_n\over M^{n-4}}\phi^n\Bigg]
\label{Sbgd}
\end{eqnarray}
For a range of reasons, one would not normally be tempted to suggest
that this theory is conformally invariant. The background metric is
taken to be non-dynamical; so that part is standard. However, the field
carries a mass term, plus there are a tower of higher dimension
operators suppressed by some mass scale $M$, the field does not carry
the conformal coupling, and the trace of the stress-tensor is non-zero.
Hence, we hope it is evident that this theory is not conformally invariant.

Nevertheless if one confuses \emph{redundancies} for symmetries, then
one might think that actually it does carry conformal symmetry.
To make this point, lets continue in the spirit of the authors and
introduce a pure gauge, or redundant, degree of freedom $\sigma$. We
now consider the following action
%
%e17 ###
%e17 #&#
%
\begin{eqnarray}
S &=& \int\! d^4x\sqrt{-g}\Bigg[{1\over2}g^{\mu\nu}e^{2\sigma
}\partial_\mu(e^{-\sigma}\phi)\partial_\nu(e^{-\sigma}\phi
)\nonumber\\
&&{}
-{1\over2}e^{2\sigma}m^2\phi^2-{\lambda\over4}\phi^4-\sum
_{n=6}^\infty{c_n\over M^{n-4}}e^{-(n-4)\sigma}\phi^n\Bigg]\,\,\,\,\,\,\,
\label{sigma}
\end{eqnarray}
This action is unchanged under the following set of gauge transformations
%
%e18 ###
%e18 #&#
%
\begin{eqnarray}
g_{\mu\nu}&\to& e^{2\alpha}g_{\mu\nu}\nonumber\\
\phi&\to& e^{-\alpha}\phi\nonumber\\
\sigma&\to&\sigma-\alpha
\label{GT2}
\end{eqnarray}
Hence, following the same reasoning that is used by these authors, one
would conclude that even this theory carries a conformal symmetry.
However, this is in fact nothing more than a field redefinition of
$\phi$, etc; not an actual change in the field. We can (and should)
gauge fix away this extra degree of freedom $\sigma$. We can gauge fix
$\sigma=0$ and then we recover the action in Eq.~(\ref{Sbgd}). Hence
this theory of course does \emph{not} carry conformal symmetry, even
though it can be rewritten in a way that gives the impression that it
does (for instance it is simple to check that the trace of the
stress-tensor $T^\mu_\mu$ that is derived from Eq.~(\ref{sigma}) is
non-zero). We hope this makes it very clear that pure gauge versions of
conformal symmetries are not real symmetries.

%s5 ###
%s5 #&#
\section{Global symmetries}\label{Global}

While these models do not possess conformal symmetry, they do possess a
global $SO(1,1)$ symmetry that relates $\psi$ and $\chi$. An attempt
to deform this global symmetry is presented by the introduction of the
function $F(\psi/\chi)$ in Eq.~(\ref{GIactionDeform}).
However, it
is unusual to deform a symmetry by introducing a function that depends
on a redundant degree of freedom. This inevitably means that the power
counting that is being envoked is scrambled by the redundancy. Instead
to make the symmetry and its deformations manifest, it is best to first
remove this extra redundant degree of freedom by gauge fixing to the
Einstein frame. With the symmetry in place this simplifies to
Eq.~(\ref{GFaction}) which carries a manifest global
symmetry: a shift symmetry
%
%e19 ###
%e19 #&#
%
\begin{eqnarray}
\phi\to\phi+\phi_0
\end{eqnarray}
Indeed, the Einstein frame is the frame that makes symmetries as
manifest as possible. In the next section we examine this shift
symmetry in a rigorous way.

%s6 ###
%s6 #&#
\section{Effective field theory}\label{EFT}

So, having gauge fixed to the Einstein frame, to make the symmetries
manifest, we can begin deforming away from this shift symmetry. There
are two basic ways to do this: (i) perturbatively, and (ii) non-perturbatively.
In this section we will describe how to deform the symmetries in a
systematic and controlled way, according to the principles of effective
field theory.

Firstly, we note that the starting action that carries the
shift-symmetry (Eq.~(\ref{GFaction})) is non-renormalizable. There
will inevitably be an infinite tower of corrections to the action.
However, the corrections that are generated perturbatively will respect
the global shift symmetry. This means that the generated corrections
will be \emph{derivative corrections}. This means the full Lagrangian
should include a tower of corrections of the form
%
%e20 ###
%e20 #&#
%
\begin{eqnarray}
\Delta\mathcal{L} = \sum_{n=2}{d_n \over M^{4n-4}}(\partial\phi
)^{2n}+\sum_{n=2}{g_n\over M^{4n-4}}(\mpl^2R)^{n}+\ldots
\end{eqnarray}
where the second term is shorthand for various possible contractions of
the Riemann tensor. The dots indicate various other corrections
involving higher derivative terms (box operator, etc.) and cross terms
between derivative of $\phi$ and the Riemann curvature tensor. We
cannot know what is the characteristic value of $M$, the mass scale
that sets this expansion. It would be associated with heavy fields that
we integrate out. But we can, as a model building assumption, take it
to be very large, say, $M\sim\mpl$. In this case we can safely ignore
\emph{all} these higher order derivative corrections. This is because
the characteristic length scale during inflation is~$H^{-1}$, which is
several orders of magnitude longer than the Planck length, suppressing
such higher derivative terms. This means that we can simply focus on
the action in Eq.~(\ref{GFaction}), under the assumption that
$M$ is
sufficiently large, and consider how to deform the shift symmetry.

%s6.1 ###
%s6.1 #&#
\subsection{Perturbative corrections}

Let us now consider adding corrections that break the shift symmetry,
giving rise to a potential function $V(\phi)$.
For example, the first natural terms to consider is a mass term and a
possible quartic term for the classical potential
%
%e21 ###
%e21 #&#
%
\begin{eqnarray}
V_{cl}(\phi)={1\over2}m^2\phi^2+{\lambda\over4}\phi^4
\label{Vcl}
\end{eqnarray}
In order for this model to give rise to the correct amplitude of scalar
fluctuations, requires $m\lesssim10^{13}~\mbox{GeV}$, $\lambda\lesssim10^{-12}$.
Having broken the shift symmetry, one should expect a tower of
corrections to be generated at the quantum level. Indeed, graviton loops
will generate such corrections of the form
%
%e22 ###
%e22 #&#
%
\begin{eqnarray}
\Delta V=\sum_{n=6}{c_n\over\mpl^{n-4}}\phi^n
\end{eqnarray}
However, it is very important to note the role of symmetry. Since the
shift symmetry is restored in the $m,\lambda\to0$ limit, then so too
should these quantum general corrections. Indeed, at one loop, one finds
that the quantum generated corrections to a classical potential take
the form
%
%e23 ###
%e23 #&#
%
\begin{eqnarray}
\Delta V_{1\mbox{\tiny{loop}}}= \left(a_1{V_{cl}''(\phi
)V_{cl}(\phi)\over(4\pi
)^2\mpl^2}+a_2{V_{cl}^2(\phi)\over(4\pi)^2\mpl^4}\right)\log
(\phi)\,\,\,\,\,\,\,
\end{eqnarray}
where $a_{1,2}=\mathcal{O}(1)$ numbers that do not concern us here.
Evaluating this for $m\ll\mpl$, $\lambda\ll1$, and $\phi\sim\mpl
$, we see that these corrections are negligibly small. Hence the
classical potential in (\ref{Vcl}) is stable against perturbative
quantum gravity corrections.
One might be concerned that it is not technically natural for the mass
to be small, but this is only a problem if the $\phi$ interactions are
sufficiently large. So if we take the limit in which we ignore $\lambda
$ at the classical level. Then the residual potential
%
%e24 ###
%e24 #&#
%
\begin{eqnarray}
V_{cl}(\phi)={1\over2}m^2\phi^2
\label{Vcl2}
\end{eqnarray}
leaves a mass whose value is technically natural to be $m\ll\mpl$ as
there are no scalar--scalar interaction to drive it to large values.
There are graviton corrections only, which are Planck suppressed,
leading to reasonably small corrections to $\Delta m^2$. Hence this
classic model of inflation \cite{Linde2} is stable against
perturbative quantum gravity corrections that arise in the effective
field theory, and the mass itself is stable against radiative corrections.
Hence, a consistent use of effective field theory around a shift
symmetry leads to a candidate simple model for inflation.$^2$\footnote
{$^2$ Recently, in Refs.~\cite{Hertzberg:2013jba,Hertzberg:2013mba} it
was found that one can go even further:
by promoting $\phi$ to a
complex field with an approximate $U(1)$ symmetry, one can also achieve
baryogenesis in this simple model of inflation organized by symmetry.}
Its cosmological predictions are
%
%e25 ###
%e25 #&#
%
\begin{eqnarray}
n_s\approx1-{2\over N_e},\,\,\,\,\,r\approx{8\over N_e}
\end{eqnarray}
which we will discuss further in Section~\ref{Bmode}.

Note this does not mean that this model will be readily attainable in a
top-down approach. That is, it is non-trivial to obtain this low energy
effective field theory from a microscopic theory. One needs to obtain the appropriate mass scale and the approximate shift symmetry to be respected to an excellent accuracy.
The reason this is not trivial to achieve is that the field value $\phi$ is super-Planckian during inflation. By computing the evolution of the field during the course of inflation, it is simple to show
%
%e26 ###
%e26 #&#
%
\begin{eqnarray}
\Delta\phi\approx2\sqrt{N_e}\,\mpl
\end{eqnarray}
A microscopic theory may give rise to a large tower of Planck suppressed corrections even at the level of the classical effective potential.
So although this low energy Lagrangian is radiatively stable, it is
unclear if it will arise from a microscopic theory. 

One way to potentially avoid the super-Planckian behavior of $\phi$ is
to consider a large number of fields; this appears in the so-called
``N-flation'' models \cite{Dimopoulos:2005ac}. One can check that for a
typical field $\phi_i$, its typical displacement is (using the
Pythagorean theorem)
%
%e27 ###
%e27 #&#
%
\begin{eqnarray}
\Delta\phi\approx2\sqrt{N_e}\,\mpl/\sqrt{N}
\end{eqnarray}
where $N$ is the number of scalars. For $N$ of a few hundred, this
leads to sub-Planckian field values. This is helpful in gaining control
over various higher order corrections that naturally emerge in top-down
models. Although it is not clear if all corrections can be kept under
control in string compactifications.

%s6.2 ###
%s6.2 #&#
\subsection{Non-perturbative corrections}\label{nonPert}

Another possible way to deform around the shift symmetry, is to note
that all global symmetries are expected to be broken in quantum
gravity. This does not necessarily imply a perturbative breaking, but a
possible non-perturbative breaking of the shift symmetry. Or it may be
broken by some other type of non-perturbative dynamics.

For definiteness, imagine that $\phi$ is a Goldstone boson associated
with the spontaneous breaking of a global symmetry. In this case, the
field is must be periodic. Lets call the symmetry breaking scale $F$,
leading to a period $\phi_{\mbox{\tiny{period}}}=2\pi F$. In this case the
non-perturbative generated corrections must be a collection of
harmonics of the form
%
%e28 ###
%e28 #&#
%
\begin{eqnarray}
V=V_0+\sum_{n=1}V_n\cos(n\phi/F)
\end{eqnarray}
The coefficients $V_n$ may be associated with some non-perturbative
effect, such as instantons.
In some cases, we can imagine that the leading harmonic is dominant. So
lets approximate the potential as a single cosine. By setting aside the
(late-time) cosmological constant, we write the potential as
%
%e29 ###
%e29 #&#
%
\begin{eqnarray}
V={V_0\over2}\left(1+\cos(\phi/F)\right)
\label{Vcosine}
\end{eqnarray}
This is the so-called ``natural inflation'' model \cite{Adams:1992bn}.
For details of the predictions for $n_s$ and $r$, see the Appendix. We
will discuss this further in Section~\ref{Bmode}.
One finds that in order to achieve a nearly scale invariant spectrum,
the parameter $F$ must satisfy $F\gtrsim\mpl$.
This does not seem trivial to achieve, as it would indicate a
super-Planckian symmetry breaking scale. A related direction is to
imagine a field $\phi$ that moves in some ``spiral'' in field space,
via a so-called ``monodromy'' \cite
{Silverstein:2008sg,McAllister:2008hb}; these models also seem promising.

%s7 ###
%s7 #&#
\section{Consequences for B-modes}\label{Bmode}

Here we examine the consequences for the amplitude of primordial
B-modes that arise from the tensor modes generated during inflation. We
will consider two different classes of large-field models: namely those
built on a cancellation of terms that tend to appear in the
``conformal'' models and elsewhere in the literature, and those built on
deforming around a shift symmetry. We will then also consider small
field models.

%s7.1 ###
%s7.1 #&#
\subsection{Models based on fine tuning}

There exist many large field models ($\Delta\phi\gtrsim\mpl$) that
rely upon the cancellation of a tower of terms in the potential. For
instance, lets return to the ``conformal symmetry'' models described
earlier in the paper (recall that they do not carry a real conformal
symmetry, but only a redundancy). Recall that the potential in the
Einstein frame took the form $V\sim F(\tanh(\phi/\sqrt{6}\,\mpl))$.
For some simple choices of the function $F$, this leads to models that
at large field values take the form
%
%e30 ###
%e30 #&#
%
\begin{eqnarray}
V(\phi)\approx V_0\left(1-e^{-\sqrt{2\over3}{\phi\over\mpl
}}\right)
\label{Vexp}
\end{eqnarray}
(we have absorbed a possible overall coefficient of the exponential
into $\phi$.) This leads to the tensor-to-scalar ratio, that we
mentioned earlier, of $r\approx12/N_e^2$. There are various types of
models that tend to this exponentially flat behavior at large $\phi$
(including the original $R+R^2$ model \cite{Starobinsky}, large
non-minimal coupling models \cite{Salopek:1988qh,Bezrukov:2007ep},
etc). For $N_e\sim55$, this leads to $r\approx0.003$. This is
consistent with WMAP and Planck data \cite
{Hinshaw:2012aka,Ade:2013uln}, 
and will require significant improvement in technology to detect (including the identification of various foregrounds that can contaminate B-modes)

However, as we showed earlier, these models arise from not rigorously
deforming around a manifest symmetry according to the principles of
effective field theory. This can be seen here in this result for the
potential $V(\phi)$. The potential is a tower of operators in powers
of $\phi$. This tower has the amazing property that the terms tend to
cancel against one another at large $\phi$, so as to produce an
asymptotically flat $V$.
Another way to see this is to introduce the $SO(1,1)$ breaking term $F$
in a different way, such as
%
%e31 ###
%e31 #&#
%
\begin{eqnarray}
V_J(\psi,\chi)={\lambda\over4}(\psi^2-F(\psi/\chi)\chi^2)^2
\end{eqnarray}
As long as $F(\psi/\chi)\neq1$ for large $\psi$, $\chi$, then this
special flatness does not occur. For example, if we choose $F=1.01$,
this would appear to be some ``small'' breaking of the $SO(1,1)$
symmetry, but it ruins the asymptotic flatness.
Instead one is typically lead to completely different potentials in the
Einstein frame.

So since the coefficients in the above (\ref{Vexp})
exponential for
$V$ are \emph{not} determined by symmetry (recall that the underlying
theory does not carry any conformal symmetry, and the global symmetry
was scrambled when the action was formulated) this is a form of \emph
{fine tuning}. (This effects other models also \cite
{Barbon:2009ya,Hertzberg:2011rc}.) The coefficients are chosen to
reproduce this special function, even though there is no symmetry that
actually organizes them into this form. One consequence of this very
special choice of coefficients, leading to this very special
exponentially flat potential, is that the amplitude of B-modes is
small.$^3$\footnote{$^3$ In other contexts, one can certainly
introduce other sorts of fine tunings to obtain large amplitude B-modes also.}

%s7.2 ###
%s7.2 #&#
\subsection{Models based on symmetry}

On the other hand, by expanding around a shift symmetry according to
the principles of effective field theory, it is more common to produce
potentials that continue to change at large field values, rather than
flatten to a constant. As we mentioned earlier, if we introduce a mass
term as the leading term that breaks the shift symmetry $V(\phi
)={1\over2}m^2\phi^2+\ldots$\,, we will not generate large corrections
within the effective field theory. Furthermore, this leads to a
consistent large field model of inflation that does not rely upon a
tower of operators whose coefficients conspire to cancel against one
another. Instead, higher corrections, such as $\lambda\phi^4$, tend
to steepen the potential.

Furthermore, if one has some knowledge of the microscopic theory, one
might be led to other sorts of potentials. For example a periodic
potential would naturally emerge for a Goldstone boson that arises from
a symmetry that is broken by non-perturbative quantum effects. Other
possibilities include fields whose shift symmetry is maintained,
approximately, by a monodromy over large field ranges.

In these types of models, there is no general preference for the field
to become asymptotically flat. Rather the symmetry may simply protect
the potential to remain ``sufficiently flat'' over large field values
for inflation to occur. Generally this leads to relatively large
amplitude B-modes. For instance, in the $V\sim m^2\phi^2$ model, the
prediction of $r\approx8/N_e$ leads to $r\approx0.15$. For monodromy
models, the predictions are comparable, though a little smaller. For
the case of the cosine potential, arising from non-perturbative quantum
effects, the prediction is $r\leq0.15$, depending on the ratio $F/\mpl
$. In general, these amplitudes for B-modes should be detectable in upcoming CMB experiments, although it is unclear
if they are completely compatible with existing Planck data \cite{Ade:2013uln}.

%s7.3 ###
%s7.3 #&#
\subsection{Small field models}

Another possibility is to focus on small field models. In this case, a
tower of corrections suppressed by the Planck scale seems less
problematic. However, one should at least be concerned about the $\sim
\phi^6/\mpl^2$ term from spoiling the flatness of the potential. This
is sometimes referred to the $\eta$-problem. This quintic piece can
raise $\eta$, leading to only a small number of e-foldings of
inflation. So in this case, one only needs to fine tune a single
operator to be small, which seems more reasonable.

These models are constrained to produce negligible gravity waves, or
B-modes in the CMB, by the ``Lyth bound'' \cite{Lyth:1996im}
%
%e32 ###
%e32 #&#
%
\begin{eqnarray}
r < 0.5\left({\Delta\phi\over\mpl}\right)^2
\end{eqnarray}
So for reasonably large values of $r$, namely $r\gtrsim0.1$, these
small field models are not allowed as $\Delta\phi$ would need to be
of the order of or greater than $\mpl$. Such models would be ruled out by a discovery of B-modes.

%s8 ###
%s8 #&#
\section{Discussion}\label{Discuss}

%s8.1 ###
%s8.1 #&#
\subsection{Could gravity be conformal?}\label{ConfGravity}

Earlier we examined the claims in the literature that standard Einstein
gravity with a cosmological constant is in fact a conformal field
theory. We showed that in fact this theory does not carry conformal
symmetry, instead authors were introducing only a redundancy into the
description.
However, it is interesting to examine whether some substantial
modifications to standard Einstein gravity might actually result in a
conformal theory.

One interesting possibility is that the Newton's constant flows at high
energies to a fixed point due to quantum corrections \cite{Weinberg}.
In addition, one would need all couplings to flow to a fixed point (and
there would be infinitely many). In this case the theory would flow to
a conformal field theory in the UV.
This is interesting to pursue, but may be incompatible with the density
of states of black holes \cite{Shomer:2007vq}. Instead, the counting
of states in the UV for black holes is comparable to the counting of
states of a conformal field theory in one lower dimension. This is
related to the famous AdS/CFT correspondence.
Another possible way that gravity could be conformal is to consider
Weyl gravity and its variants (although it is unclear if such theories
can be made sensible).

%s8.2 ###
%s8.2 #&#
\subsection{Effective field theory and quantum gravity}

We showed that a useful way to build simple models of inflation is to
start with a shift symmetry for a scalar field and deform around it.
From the effective field theory, this is a consistent approach as it
leads to models that are radiatively stable; the perturbatively
generated quantum corrections are small. We showed that simple models,
including either perturbative or non-perturbative corrections, tend to
lead to slowly varying potentials, without fine tuning, and typically
large B-modes.

It is important to note that these models lead to large, typically
super-Planckian field excursions. The Hubble scale being probed is well
below the Planck energy, so the effective field theory is consistent,
but it is obviously sensitive to the details of the UV completion. So
it is of great importance to embed inflation within quantum gravity to
obtain full control over these higher dimension operators in the
effective potential. In other words, it is important to check if these
simple symmetry arguments persist in the full quantum gravity theory,
or if important modifications are present.

Observational data, including the possibility of a positive detection
of B-modes, is very important to address these questions.

\begin{center}
{\bf Acknowledgments}
\end{center}

We would like to acknowledge support by the Center for Theoretical Physics at MIT. 
This work is supported by the U.S. Department of Energy under cooperative research agreement Contract Number DE-FG02-05ER41360.

\section*{Appendix}\label{appA}

In this appendix, we describe the predictions for the spectral index
$n_s$ and tensor-to-scalar ratio in simple single field models and then
apply the analysis to the cosine potential of Section~\ref{nonPert}.

The spectral index $n_s$ and the tensor-to-scalar ratio $r$ are related
to the slow-roll parameters $\epsilon$ and $\eta$ by the following formulas
%
%e33 ###
%e33 #&#
%
\begin{eqnarray}
n_s = 1- 6\epsilon_*+2\eta_*,\,\,\,\,\,\,r=16\epsilon_*
\end{eqnarray}
The slow roll parameters were defined in terms of derivatives of the
potential $V$ in Eq.~(\ref{slowroll}). The * subscript here indicates
that they need to be evaluated at the special moment when the modes
leaves that we are interested in (namely those that affect the CMB).
This is usually expressed in terms of the number of e-foldings of
inflation $N_e$, which is given by
%
%e34 ###
%e34 #&#
%
\begin{eqnarray}
N_e = {1\over\mpl}\int^{\phi{_*}}_{\phi_{e}}\!\!{d\phi\over\sqrt
{2\,\epsilon(\phi)}}
\end{eqnarray}
($\phi_e$ is the end of inflation).

In the case of the cosine potential given in Eq.~(\ref{Vcosine}), we find
%
%e36 ###
%e35 ###
%e35 #&#
%e36 #&#
%
\begin{eqnarray}
\epsilon_* &=& {\mpl^2\over2 F^2}\tan^2(\phi_*/2F)\\
\eta_* &=& -{\mpl^2\over F^2}{\cos(\phi_*/F)\over1+\cos(\phi_*/F)}
\end{eqnarray}
and the number of e-foldings is given by
%
%e37 ###
%e37 #&#
%
\begin{eqnarray}
N_e={2F^2\over\mpl^2}\ln\left({\sin(\phi_e/2F)\over\sin(\phi
_*/2F)}\right)
\end{eqnarray}
This allow a parametric representation of $n_s$ and $r$ as we vary the
dimensionless quantity $F/\mpl$ for a given choice of $N_e$.
For $F\gg\mpl$ it is simple to show that this reproduces the
predictions of $V\sim m^2\phi^2$, including a near scale invariant
spectrum. On the other hand, as we decrease $F$ below $\mpl$, the
predictions deviate from scale invariance more and the tensor to scalar
ratio decreases.
%\end{appm}

\end{document}